\documentclass{article}
\usepackage{amsmath}

\begin{document}

\begin{center}
\bigskip

\textbf{FLOW OF THE VISCOUS-ELASTIC LIQUID IN THE }

\textbf{NON- HOMOGENEOUS TUBE }

\bigskip

\textbf{V.YU. BABANLY}
\begin{center}
{Institute of Applied Mathematics, Baku State University}
\end{center}
{Z. Khalilov 23, AZ 1148, Baku, Azerbaijan}
\end{center}

\bigskip

\begin{quote}
\textbf{ABSTRACT.}{\small \ }\textit{A problem on propagation of
waves in deformable shells with flowing liquid is very urgent in
connection with wide use of liquid transportation systems in living
organisms and technology. It is necessary to consider shell motion
equations for influence of moving liquid in cavity on the dynamics
of a shell by solving such kind problems.}

\textit{Nowadays a totality of such problems is a widely developed
field of hydrodynamics. However, a number of peculiarities connected
with taking into account viscous-elastic properties of the liquid
and inhomogeneity of the shell material generates considerable
mathematical difficulties connected with integration of boundary
value problems with variable coefficients.}

\textit{In the paper we consider wave flow of the liquid enclosed in
deformable tube. The used mathematical model is described by the
equation of motion of incompressible viscous elastic liquid combined
with equation of continuity and dynamics equation for a tube
inhomogeneous in length. It is accepted that the tube is cylindric,
semi-infinite and rigidly fastened to the environment. At the
infinity the tube is homogeneous. As a final result, the problem is
reduced to the solution of Volterra type integral equation that is
solved by sequential approximations method. Pulsating pressure is
given at the end of the tube to determine the desired hydrodynamic functions.%
}

KEYWODS. viscous-elastic liquid , non- homogeneous tube{\small \ }
\end{quote}

{\small \bigskip }

Statement and mathematical ground of hydroelasticity problem is
considered that admits to describe small amplitude wave propagation
in elastic tube nonhomogeneous in length (with modulus of elasticity
$E=E\left( x\right) $ and density $\rho _{m}=\rho _{m}\left(
x\right) $), containing liquid. The basis of liquid's model is the
accounting of its viscous-elastic properties. Here, one-dimensional
linearized equations are used.

1. Given a rectilinear semi-infinite cylindrical thin-shelled tube
of constant thickness $h$ and radius $R$ whose material is subjected
to Hooke's law. In the considered case the system of hydroelasticity
equations is of the form [1,2]
\begin{equation*}
\frac{\partial u\left( x,t\right) }{\partial
x}+\frac{2}{R}\frac{\partial w\left( x,t\right) }{\partial
t}=0;\eqno(1)
\end{equation*}

\begin{equation*}
\rho _{f}\frac{\partial u\left( x,t\right) }{\partial t}=\frac{\partial }{%
\partial x}\left\{ \sigma \left( x,t\right) -p\left( x,t\right) \right\} ;%
\eqno(2)
\end{equation*}

\begin{equation*}
\prod\limits_{j=1}^{r}\left\{ \sigma \left( x,t\right) +\lambda _{j}\frac{%
\partial \sigma \left( x,t\right) }{\partial t}\right\} =2\eta
\prod\limits_{j=1}^{s}\left\{ \frac{\partial u\left( x,t\right) }{\partial x}%
+\theta _{j}\frac{\partial ^{2}u\left( x,t\right) }{\partial x\partial t}%
\right\} ;\eqno(3)
\end{equation*}

\begin{equation*}
p\left( x,t\right) -\frac{h}{R^{2}}E\left( x\right) w\left(
x,t\right) =h\rho _{m}\left( x\right) \frac{\partial ^{2}w\left(
x,t\right) }{\partial t^{2}}.\eqno(4)
\end{equation*}

Rheological relation (3) describes sufficiently well the flow of
liquid continua containing long \textquotedblleft
high-molecular\textquotedblright\ compounds. It should be noted that
there exist two classes of variants of model (3). The media
possessing instantaneous elasticity for which $r=s+1$ belong to the
first class. The models detecting viscous behavior at instantaneous
loading belong to the second class. For them $r=s$.

At the equations (1) - (4) $u\left( x,t\right) $ is longitudinal
speed of flow of liquid, $w\left( x,t\right) $ is radial
displacement of tube's walls, $p\left( x,t\right) $ is pressure,
$\sigma \left( x,t\right) $ is \textquotedblleft
viscous-elastic\textquotedblright\ stress, $\rho _{f}$ is liquid's
density, $\eta $ is its dynamic viscosity coefficient, the
quantities $\lambda _{j}$ and $\theta _{j}$ form relaxation and
retardation spectra, respectively.

Not loosing generality, we represent the functions $E\left( x\right) $ and $%
\rho _{m}$ by means of the equalities $E\left( x\right) =E_{\infty
}g_{1}\left( x\right) $ and $\rho _{m}\left( x\right) =\rho
_{m\infty }g_{2}\left( x\right) $ and accept that the functions
$g_{1}\left( x\right) $ and $g_{2}\left( x\right) $ are twice
differentiable. We'll also assume that the tube is homogeneous at
infinity. Hence we have:

\begin{equation*}
\underset{x\rightarrow \infty }{\lim }g_{1}\left( x\right) =\underset{%
x\rightarrow \infty }{\lim }g_{2}\left( x\right) =1.\eqno(5)
\end{equation*}

At the same time we assume
\begin{equation*}
\underset{x\rightarrow \infty }{\lim }g_{1}^{\prime }\left( x\right) =0;~%
\underset{x\rightarrow \infty }{\lim }g_{2}^{\prime }\left( x\right)
=0;~\ \underset{x\rightarrow \infty }{\lim }g_{1}^{\prime \prime
}\left( x\right) =0;~\ \underset{x\rightarrow \infty }{\lim
}g_{2}^{\prime \prime }\left( x\right) =0,\eqno(6)
\end{equation*}%
where here and later on the primes mean differentiation with respect
to $x$.

Transforming the system (1)-(4) by formula

\begin{equation*}
Q\left( x,t\right) =\pi R^{2}u\left( x,t\right)
\end{equation*}%
we get the following expance system of equation

\begin{equation*}
\frac{\partial Q\left( x,t\right) }{\partial x}+2\pi R\frac{\partial
w\left( x,t\right) }{\partial x}=0,\eqno(7)
\end{equation*}

\begin{equation*}
\rho _{f}\frac{\partial Q\left( x,t\right) }{\partial t}=\pi R^{2}\frac{%
\partial }{\partial x}\left\{ \sigma \left( x,t\right) -p\left( x,t\right)
\right\} ,\eqno(8)
\end{equation*}

\begin{equation*}
\pi R^{2}\prod\limits_{j=1}^{r}\left\{ \sigma \left( x,t\right) +\lambda _{j}%
\frac{\partial \sigma \left( x,t\right) }{\partial t}\right\} =2\eta
\prod\limits_{j=1}^{s}\left\{ \frac{\partial Q\left( x,t\right) }{\partial x}%
+\theta _{j}\frac{\partial ^{2}Q\left( x,t\right) }{\partial t\partial x}%
\right\} ,\eqno(9)
\end{equation*}

\begin{equation*}
p\left( x,t\right) -\frac{hE_{\infty }}{R^{2}}g_{1}\left( x\right)
w\left( x,t\right) =\rho _{m\infty }hg_{2}\left( x\right)
\frac{\partial ^{2}w\left( x,t\right) }{\partial t^{2}}.\eqno(10)
\end{equation*}

We separate variables and look for the solution of system (1) - (4)
in the following form:

\begin{equation*}
\left.
\begin{array}{l}
Q\left( x,t\right) =Q_{1}\left( x\right) \exp \left( i\omega
t\right) ,~\
~w\left( x,t\right) =w_{1}\left( x\right) \exp \left( i\omega t\right) , \\
\sigma \left( x,t\right) =\sigma _{1}\left( x\right) \exp \left(
i\omega t\right) ,~\ ~p\left( x,t\right) =p_{1}\left( x\right) \exp
\left( i\omega
t\right) ,%
\end{array}%
\right. \eqno(11)
\end{equation*}%
that allows to reduce the initial system of equations to the system
of differential equations of the form

\begin{equation*}
Q_{1}^{1}\left( x\right) +2\pi R\omega iw_{1}\left( x\right)
=0,\eqno(12)
\end{equation*}

\begin{equation*}
\rho _{f}i\omega Q_{1}\left( x\right) =\pi R^{2}\left( \sigma
_{1}^{1}\left( x\right) -p_{1}^{1}\left( x\right) \right) ,\eqno(13)
\end{equation*}

\begin{equation*}
p_{1}\left( x\right) =\left\{ \frac{hE_{\infty }}{R^{2}}g_{1}\left(
x\right) -h\omega ^{2}\rho _{m\infty }g_{2}\left( x\right) \right\}
w_{1}\left( x\right) ,\eqno(14)
\end{equation*}

\begin{equation*}
\sigma _{1}\left( x\right) =\frac{2\eta }{\rho R^{2}}\frac{b}{a}%
Q_{1}^{1}\left( x\right) ,\eqno(15)
\end{equation*}%
where in

\begin{equation*}
\prod\limits_{j=1}^{r}\left( 1+i\omega \lambda _{j}\right)
=a=a_{0}+ia_{1},~\ \ \prod\limits_{j=1}^{s}\left( 1+i\omega \theta
_{j}\right) =b=b_{0}+ib_{1}
\end{equation*}%
is accepted, and $\omega $ is the given angular frequency. Further
we introduce the denotation

\begin{equation*}
G\left( x\right) =\frac{2\eta }{\rho _{f}}\omega i\frac{b}{a}-\frac{c_{0}^{2}%
}{\omega ^{2}}g_{1}\left( x\right) +\frac{Rh\rho _{m\infty }}{2\rho _{f}}%
g_{2}\left( x\right) ,~~\left( c_{0}=\frac{hE_{\infty }}{2R\rho
_{f}}\right) ,\eqno(16)
\end{equation*}%
whence

\begin{equation*}
G^{\prime }\left( x\right) =\frac{Rh\rho _{m\infty }}{2\rho _{f}}%
g_{2}^{\prime }\left( x\right) -\frac{c_{0}^{2}}{\omega
^{2}}g_{1}^{\prime }\left( x\right) .\eqno(17)
\end{equation*}

. (17)

Combining equations (12-15), after some transformations we get the
following equation for the function\ $u_{1}\left( x\right) $

\begin{equation*}
G\left( x\right) u_{1}^{\prime \prime }\left( x\right) +G^{\prime
}\left( x\right) u_{1}^{\prime }\left( x\right) -u_{1}\left(
x\right) =0.\eqno(18)
\end{equation*}

We use the Liouville substitution

\begin{equation*}
y\left( x\right) =Q_{1}\left( x\right) \exp \left\{ \frac{1}{2}\int \frac{%
G^{\prime }\left( x\right) }{G\left( x\right) }dx\right\}
=Q_{1}\left( x\right) \sqrt{\left\vert G\left( x\right) \right\vert
}\eqno(19)
\end{equation*}%
and reduce equation (18) to the form

\begin{equation*}
y^{\prime \prime }\left( x\right) +I\left( x\right) y\left( x\right) =0,\eqno%
(20)
\end{equation*}%
where the invariant $I\left( x\right) $ is determined by the
relation

\begin{equation*}
I\left( x\right) =\frac{1}{4}\left\{ \frac{G^{\prime }\left( x\right) }{%
G\left( x\right) }\right\} ^{2}-\frac{1}{2}\frac{G^{\prime \prime
}\left( x\right) }{G\left( x\right) }-\frac{1}{G\left( x\right)
}.\eqno(21)
\end{equation*}

Using relations (5) and (6) we establish the following limit
equalities:

\begin{equation*}
\underset{x\rightarrow \infty }{\lim }G\left( x\right) =\frac{2\eta
}{\rho _{f}\omega }\frac{b}{ia}-\frac{c_{0}^{2}}{\omega
^{2}}+\frac{Rh\rho _{m\infty }}{2\rho _{f}},~~\
\underset{x\rightarrow \infty }{\lim }G^{\prime }\left( x\right) =0.
\end{equation*}

Hence we follow (21) and get the expression

\begin{equation*}
\underset{x\rightarrow \infty }{\lim }I\left( x\right) =\delta ^{2}=\frac{%
k_{0}}{k_{0}^{2}+k_{1}^{2}}-i\frac{k_{1}}{k_{0}^{2}+k_{1}^{2}},\eqno(22)
\end{equation*}%
wherein

\begin{equation*}
k_{0}=\frac{2\eta }{\rho _{f}\omega }\frac{a_{1}b_{0}-a_{0}b_{1}}{%
a_{0}^{2}+a_{1}^{2}}+\frac{c_{0}^{2}}{\omega ^{2}}+\frac{Rh\rho _{m\infty }}{%
2\rho _{f}}
\end{equation*}

\begin{equation*}
k_{1}=\frac{2\eta }{\rho _{f}\omega }\frac{a_{0}b_{0}-a_{1}b_{1}}{%
a_{0}^{2}+a_{1}^{2}}.
\end{equation*}

Now, using denotation

\begin{equation*}
q\left( x\right) =1-\frac{I\left( x\right) }{\delta ^{2}},\eqno(23)
\end{equation*}%
we reduce equation (20) to the form

\begin{equation*}
y^{\prime \prime }\left( x\right) +\delta ^{2}y\left( x\right)
=\delta ^{2}q\left( x\right) y\left( x\right) .\eqno(24)
\end{equation*}

We'll use a root at which ${Im} \delta<0$ later on, and on the
potential $q\left( x\right)$ we impose the integrability condition

\begin{equation*}
\int\limits_{0}^{\infty }\left\vert q\left( x\right) \right\vert dx<+\infty .%
\eqno(25)
\end{equation*}

In order to construct the solutions, equation (24) should be
completed with the following boundary conditions

\begin{equation*}
y\left( 0\right) =y_{0},~\ y\rightarrow 0\text{ as }x\rightarrow \infty .%
\eqno(26)
\end{equation*}

The quantity $y_{0}$ depends on the functional rejime of the system,
and the second condition (26) provides the boundedness of the
desired solution. As a result, the solution of the hydroelasticity
problem is reduced to a singular boundary value problem of
Sturm-Lioville type (24) and (26) under the condition (25).

\textbf{2.} In equation (24) considering $\delta ^{2}q\left(
x\right) $ as an external source and applying the method of
arbitrary constants variation we can reduce the solution of the
problem to the equivalent integral equation [3]

\begin{equation*}
y\left( x,-\delta \right) =Ce^{-i\delta x}+\delta
\int\limits_{x}^{\infty }\sin \delta \left( \xi -x\right) q\left(
\xi \right) y\left( \xi ,-\delta \right) d\xi ,\eqno(27)
\end{equation*}%
Here the constant $C$ is determined as

\begin{equation*}
C=\frac{y_{0}}{f\left( 0,-\delta \right) ,}
\end{equation*}%
and

\begin{equation*}
y=y_{0}\frac{f\left( x,-\delta \right) }{f\left( 0,-\delta \right)
},
\end{equation*}%
where by [4] a new function $f\left( x,-\delta \right) $ is
determined from the solution of the integral equation

\begin{equation*}
f\left( x,-\delta \right) =e^{-i\delta x}+\delta
\int\limits_{x}^{\infty }\sin \delta \left( \xi -x\right) q\left(
\xi \right) f\left( \xi ,-\delta \right) d\xi .\eqno(28)
\end{equation*}

Equation (28) is a Volterra type equation and is solved by the
sequential approximations method:

\begin{equation*}
f\left( x,-\delta \right) =\sum\limits_{n=0}^{\infty }\delta
^{n}f_{n}\left( x,-\delta \right) .\eqno(29)
\end{equation*}

Therein we have of the following recurrent relations

\begin{equation*}
\left.
\begin{array}{l}
f_{0}\left( x,-\delta \right) =e^{-i\delta x}, \\
........................... \\
f_{n}\left( x,-\delta \right) =\delta \int\limits_{x}^{\infty }\sin
\delta \left( \xi -x\right) q\left( \xi \right) f_{n-1}\left( \xi
,-\delta \right)
d\xi ~\ \ \ \left( n=1,2,...\right) .%
\end{array}%
\right. \eqno(30)
\end{equation*}

By inequality (25) from uniform convergences of sequential
approximations by the Weierstrass test we can establish that the
unique solution of integral equation (28) is determined by relation
(29). We can directly establish that this solution solution of input
equation (24) as well.

Further, it is easy to determine the functions from systems
(12)-(15) and write $u_{1}$, $w_{1}$, $p_{1}$, $\sigma _{1}$

\begin{equation*}
Q\left( x,t\right) =\frac{\pi R^{2}y\left( x\right)
}{\sqrt{\left\vert G\left( x\right) \right\vert }}y_{0}e^{i\omega
t}F\left( x\right) ;\eqno(31)
\end{equation*}

\begin{equation*}
w\left( x,t\right) =-\frac{Ry_{0}}{2\omega i}e^{i\omega t}F^{\prime
}\left( x\right) ;\eqno(32)
\end{equation*}

\begin{equation*}
\sigma \left( x,t\right) =2\eta y_{0}\frac{b}{a}e^{i\omega
t}F^{\prime }\left( x\right) ;\eqno(33)
\end{equation*}

\begin{equation*}
p\left( x,t\right) =y_{0}e^{i\omega t}\left[ \frac{hE_{\infty }i}{2R\omega }%
g_{1}\left( x\right) -\frac{Rh\rho _{m\infty }i}{2}g_{2}\left( x\right) %
\right] F^{\prime }\left( x\right) ,\eqno(34)
\end{equation*}%
where

\begin{equation*}
F\left( x\right) =\frac{1}{\sqrt{\left\vert G\left( x\right) \right\vert }}%
\frac{f\left( x,-\delta \right) }{f_{0}\left( 0,-\delta \right) }.
\end{equation*}%
To determine the quantity $y_{0}$ at the end of the tube $\left( x=0\right) $%
\ we give the pulsating pressure

\begin{equation*}
p\left( 0,t\right) =p_{0}\exp \left( i\omega t\right) .
\end{equation*}

Using equality (34) this circumstance admits immediately to
determine the quantity $y_{0}$. It is of the form:

\begin{equation*}
y_{0}=-i\frac{p_{0}}{\left\{ \dfrac{hE_{\infty }i}{2R\omega
}g_{1}\left( 0\right) -\dfrac{Rh\rho _{m\infty }}{2}g_{2}\left(
0\right) \right\} F^{\prime }\left( 0\right) }.
\end{equation*}%
Thus, the solution of the problem is complete. In conclusion note
that series (29) in combination with relations (30) gives
constructive representation of the desired solution and real parts
(31)-(34) represent physical quantity.

\bigskip

\begin{center}
\textbf{References}
\end{center}

[1]. Wallmir A.S.\textit{\ Shells in liquid and gas flow. Problems
of hydroelasticity.} M., "Nauka", 1979, 320 pp. (Russian).

[2]. Rutkevich I.M. \textit{On propagation of small perturbations in
viscous-elastic liquid. }Prikl. Mat. i Mekh., v.34, 1970, p.41-56.
(Russian).

[3]. Kiyasbeyli E.T. \textit{Wave flow of Maxwell liquid in variable
section viscous elastic tube.} Vestnik of Baku State University,
No3, 2005, p.87- 94. (Russian).

[4]. Amenzadeh R.Yu, Kiyasbeyli E.T. \textit{Waves in the elastic
tube variable sections with the proceeding liquid.} Proceedings of
III All Russian Conference in elasticity theory with international
participation. Rostov-na-Donu, "Novaya kniga"\ 2004, p.40-43.

\end{document}